\documentclass[11pt]{article}
\usepackage{graphicx}
\usepackage{epsfig}

\newcommand{\BABARPubYear}    {04}

\newcommand{\BABARConfNumber} {013}
\newcommand{\SLACPubNumber} {10615}
\newcommand{\LANLNumber} {0408028}

\def\sss{\scriptscriptstyle}
\def\barpd{{\raise.35ex\hbox
{${\sss (}$}}--{\raise.35ex\hbox{${\sss )}$}}}
\def\dbarp{\hbox{$D^{0}$\kern-1.25em\raise1.5ex\hbox{\barpd}}}
\def\dbarpstar{\hbox{$D^{*0}$\kern-1.65em\raise1.5ex\hbox{\barpd}}}
\def\dbarpnozero{\hbox{$D$\kern-0.85em\raise1.5ex\hbox{\barpd}}}
\def\Dstarz  {\ensuremath{D^{*0}}\xspace}

\def\Dz      {\ensuremath{D^0}\xspace}
\def\piz   {\ensuremath{\pi^{0}}\xspace}

\def\mes        {\mbox{$m_{\rm ES}$}\xspace}

\def\to         {\ensuremath{\rightarrow}\xspace}

\def\Dzbpar  {\ensuremath{\Dbar^{(*)0}}\xspace}

\RequirePackage{xspace}

\usepackage{relsize}
\def\babar{\mbox{\slshape B\kern-0.1em{\smaller A}\kern-0.1em
    B\kern-0.1em{\smaller A\kern-0.2em R}}}

\def\piz   {\ensuremath{\pi^0}\xspace}

\def\Kbar  {\kern 0.2em\overline{\kern -0.2em K}{}\xspace}

\def\Kz    {\ensuremath{K^0}\xspace}
\def\Kzb   {\ensuremath{\Kbar^0}\xspace}
\def\KzKzb {\ensuremath{\Kz \kern -0.16em \Kzb}\xspace}
\def\Kp    {\ensuremath{K^+}\xspace}
\def\Km    {\ensuremath{K^-}\xspace}

\def\KpKm  {\ensuremath{\Kp \kern -0.16em \Km}\xspace}

\def\Dbar    {\kern 0.2em\overline{\kern -0.2em D}{}\xspace}

\def\Dz      {\ensuremath{D^0}\xspace}
\def\Dzb     {\ensuremath{\Dbar^0}\xspace}
\def\DzDzb   {\ensuremath{\Dz {\kern -0.16em \Dzb}}\xspace}
\def\Dp      {\ensuremath{D^+}\xspace}
\def\Dm      {\ensuremath{D^-}\xspace}

\def\DpDm    {\ensuremath{\Dp {\kern -0.16em \Dm}}\xspace}
\def\Dstar   {\ensuremath{D^*}\xspace}
\def\Dstarb  {\ensuremath{\Dbar^*}\xspace}
\def\Dstarz  {\ensuremath{D^{*0}}\xspace}
\def\Dstarzb {\ensuremath{\Dbar^{*0}}\xspace}

\def\Bbar    {\kern 0.18em\overline{\kern -0.18em B}{}\xspace}

\def\Bz      {\ensuremath{B^0}\xspace}
\def\Bzb     {\ensuremath{\Bbar^0}\xspace}
\def\BzBzb   {\ensuremath{\Bz {\kern -0.16em \Bzb}}\xspace}
\def\Bu      {\ensuremath{B^+}\xspace}
\def\Bub     {\ensuremath{B^-}\xspace}

\def\BpBm    {\ensuremath{\Bu {\kern -0.16em \Bub}}\xspace}

\def\BorBbar    {\kern 0.18em\optbar{\kern -0.18em B}{}\xspace}
\def\DorDbar    {\kern 0.18em\optbar{\kern -0.18em D}{}\xspace}
\def\KorKbar    {\kern 0.18em\optbar{\kern -0.18em K}{}\xspace}

\mathchardef\Upsilon="7107
\def\Y#1S{\ensuremath{\Upsilon{(#1S)}}\xspace}

\def\FourS {\Y4S}

\mathchardef\Deltares="7101
\mathchardef\Xi="7104
\mathchardef\Lambda="7103
\mathchardef\Sigma="7106
\mathchardef\Omega="710A

\def\Deltabar{\kern 0.25em\overline{\kern -0.25em \Deltares}{}\xspace}
\def\Lbar{\kern 0.2em\overline{\kern -0.2em\Lambda\kern 0.05em}\kern-0.05em{}\xspace}
\def\Sigbar{\kern 0.2em\overline{\kern -0.2em \Sigma}{}\xspace}
\def\Xibar{\kern 0.2em\overline{\kern -0.2em \Xi}{}\xspace}
\def\Obar{\kern 0.2em\overline{\kern -0.2em \Omega}{}\xspace}
\def\Nbar{\kern 0.2em\overline{\kern -0.2em N}{}\xspace}
\def\Xb{\kern 0.2em\overline{\kern -0.2em X}{}\xspace}

\def\mes        {\mbox{$m_{\rm ES}$}\xspace}

\newcommand{\tev}{\ensuremath{\mathrm{\,Te\kern -0.1em V}}\xspace}
\newcommand{\gev}{\ensuremath{\mathrm{\,Ge\kern -0.1em V}}\xspace}
\newcommand{\mev}{\ensuremath{\mathrm{\,Me\kern -0.1em V}}\xspace}
\newcommand{\kev}{\ensuremath{\mathrm{\,ke\kern -0.1em V}}\xspace}
\newcommand{\ev}{\ensuremath{\mathrm{\,e\kern -0.1em V}}\xspace}
\newcommand{\gevc}{\ensuremath{{\mathrm{\,Ge\kern -0.1em V\!/}c}}\xspace}
\newcommand{\mevc}{\ensuremath{{\mathrm{\,Me\kern -0.1em V\!/}c}}\xspace}
\newcommand{\gevcc}{\ensuremath{{\mathrm{\,Ge\kern -0.1em V\!/}c^2}}\xspace}
\newcommand{\mevcc}{\ensuremath{{\mathrm{\,Me\kern -0.1em V\!/}c^2}}\xspace}

\def\mus  {\ensuremath{\rm \,\mus}\xspace}

\def\mus        {\ensuremath{\,\mu{\rm s}}\xspace}    

\def\ra                 {\ensuremath{\rightarrow}\xspace}
\def\to                 {\ensuremath{\rightarrow}\xspace}

\def\pep2{PEP-II}
\def\BF{$B$ Factory}

\def\gsim{{~\raise.15em\hbox{$>$}\kern-.85em
          \lower.35em\hbox{$\sim$}~}\xspace}
\def\lsim{{~\raise.15em\hbox{$<$}\kern-.85em
          \lower.35em\hbox{$\sim$}~}\xspace}

\def\CP                {\ensuremath{C\!P}\xspace}

\xspace

\def\jetset74   {\mbox{\tt Jetset \hspace{-0.5em}7.\hspace{-0.2em}4}\xspace}

\long\def\inst#1{\par\nobreak\kern 4pt\nobreak
    {\it #1}\par\vskip 10pt plus 3pt minus 3pt}

\begin{document}
{\pagestyle{empty}

\begin{flushright}
\babar-CONF-\BABARPubYear/\BABARConfNumber \\
SLAC-PUB-\SLACPubNumber \\
hep-ex/\LANLNumber \\
\end{flushright}

\par\vskip 2cm

\begin{center}
 \Large Search for $b \rightarrow u$ Transitions in 
$B^- \to \dbarp~K^-$ and $B^- \to \dbarpstar~~K^-$
\end{center}
\bigskip

\begin{center}
\large The \babar\ Collaboration\\
\mbox{ }\\
\today
\end{center}
\bigskip 

\begin{center}
\large \bf Abstract
\end{center}
\noindent
We report on searches for $B^- \to \dbarp~K^-$ and $B^- \to
\dbarpstar~~K^-$, with $\dbarpstar~\to \dbarp~\pi^0$ or
$\dbarpstar~\to \dbarp~\gamma$, and $\dbarp \to K^+\pi^-$ (and charge
conjugates).  These final states, which we denote as $[K^+\pi^-]_D
K^-$ and $[K^+\pi^-]_{D^*} K^-$, can be reached through the $b \to c$
transition $B^- \to D^{(*)0}K^-$ followed by the doubly
Cabibbo-suppressed $D^0 \to K^+ \pi^-$, or through the $b \to u$
transition $B^- \to \Dzbpar K^-$ followed by the Cabibbo-favored $\Dzb
\to K^+ \pi^-$, or through interference of the two.  Our results are
based on 227 million $\FourS \to B\Bbar$ decays collected with the
\babar\ detector at SLAC. We measure the ratios
    \[
    {\cal R}_{K\pi} \equiv 
    \frac{\Gamma(B^+ \to [K^-\pi^+]_D K^+) +  \Gamma(B^- \to [K^+\pi^-]_D K^-) }
         {\Gamma(B^+ \to [K^+\pi^-]_D K^+) +  \Gamma(B^- \to [K^-\pi^+]_D K^-) } = 0.013 \pm^{0.011}_{0.009},
    \]
    \[
    {\cal R^*}_{K\pi,D\piz} \equiv 
    \frac{\Gamma(B^+ \to [K^-\pi^+]_{D^* \ra D\piz} K^+) +  \Gamma(B^- \to [K^+\pi^-]_{D^* \ra D\piz} K^-) }
         {\Gamma(B^+ \to [K^+\pi^-]_{D^* \ra D\piz} K^+) +  \Gamma(B^- \to [K^-\pi^+]_{D^* \ra D\piz} K^-) } = -0.001 \pm^{0.010}_{0.006}
    \]
and
    \[
    {\cal R^*}_{K\pi,D\gamma} \equiv 
    \frac{\Gamma(B^+ \to [K^-\pi^+]_{D^* \ra D\gamma} K^+) +  \Gamma(B^- \to [K^+\pi^-]_{D^* \ra D\gamma} K^-) }
         {\Gamma(B^+ \to [K^+\pi^-]_{D^* \ra D\gamma} K^+) +  \Gamma(B^- \to [K^-\pi^+]_{D^* \ra D\gamma} K^-) } = 0.011 \pm^{0.019}_{0.013}.
    \]

\noindent We set ${\cal R}_{K\pi} < 0.030~~{\rm (90\%~C.L.)}$ and from this limit we extract the amplitude ratio 
$r_B \equiv |A(B^- \to \Dzb K^-)/A(B^- \to \Dz K^-)| < 0.23$ at the 90\% confidence level.  This
limit is obtained with the most conservative assumptions on the values of the CKM angle $\gamma$ and
the strong phases in the $B$ and $D$ decay amplitudes.  From the measurements of
${\cal R^*}_{K\pi,D\piz}$ and ${\cal R^*}_{K\pi,D\gamma}$ we 
extract 
$r^{*2}_B \equiv |A(B^- \to \Dzb K^-)/A(B^- \to \Dz K^-)|^2 = 4.6^{+15.2}_{-7.3} \times 10^{-3}$
and $r^{*2}_B < (0.16)^2$ 
at the 90\% confidence level.

\vfill
\begin{center}

Submitted to the 32$^{\rm nd}$ International Conference on High-Energy Physics, ICHEP 04,\\
16 August---22 August 2004, Beijing, China

\end{center}

\vspace{1.0cm}
\begin{center}
{\em Stanford Linear Accelerator Center, Stanford University, 
Stanford, CA 94309} \\ \vspace{0.1cm}\hrule\vspace{0.1cm}
Work supported in part by Department of Energy contract DE-AC03-76SF00515.
\end{center}

\newpage
} 

\begin{center}
\small

The \babar\ Collaboration,
\bigskip

B.~Aubert,
R.~Barate,
D.~Boutigny,
F.~Couderc,
J.-M.~Gaillard,
A.~Hicheur,
Y.~Karyotakis,
J.~P.~Lees,
V.~Tisserand,
A.~Zghiche
\inst{Laboratoire de Physique des Particules, F-74941 Annecy-le-Vieux, France }
% ...

\end{center}\newpage

\section{INTRODUCTION}
\label{sec:Introduction}
   Following the discovery of \CP violation in $B$-meson 
   decays and the measurement of the angle $\beta$
   of the unitarity triangle~\cite{cpv} associated with
   the Cabibbo-Kobayashi-Maskawa (CKM) quark mixing matrix, focus has turned
   towards the measurements of the other angles $\alpha$ and $\gamma$.
   The angle $\gamma$ is ${\rm arg}(-V_{ub}^*V^{}_{ud}/V_{cb}^*V^{}_{cd})$,
   where $V^{}_{ij}$ are CKM matrix elements;
   in the Wolfenstein convention~\cite{wolfenstein},
   $\gamma = {\rm arg}(V_{ub}^*)$.

\begin{figure}[hb]
   \begin{center}
      \epsfig{file=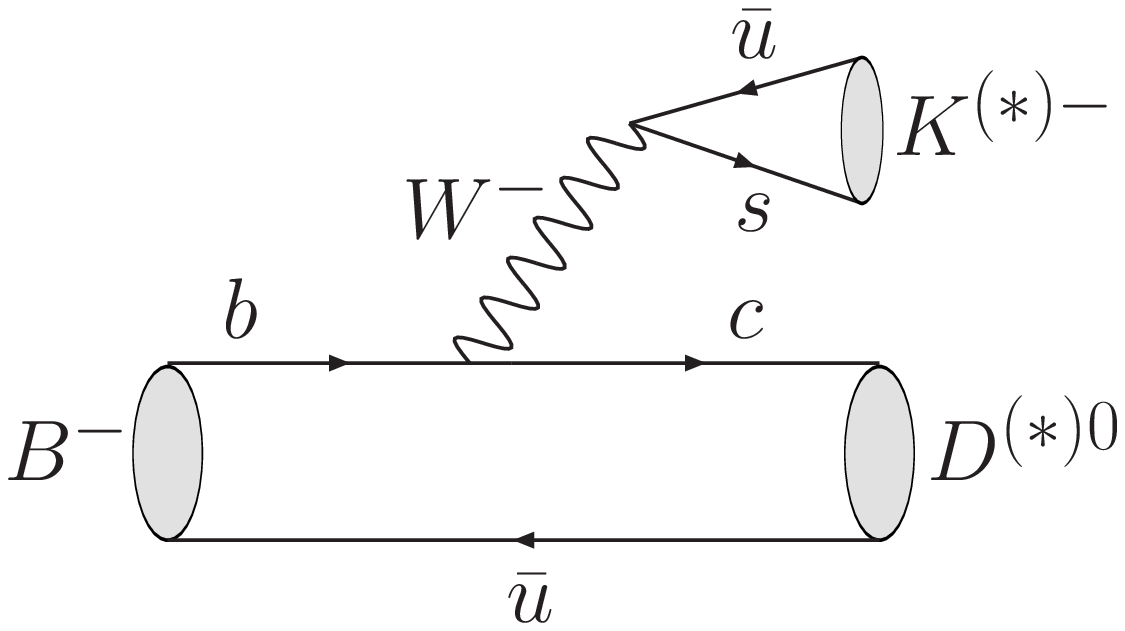,height=3.9cm}
      \epsfig{file=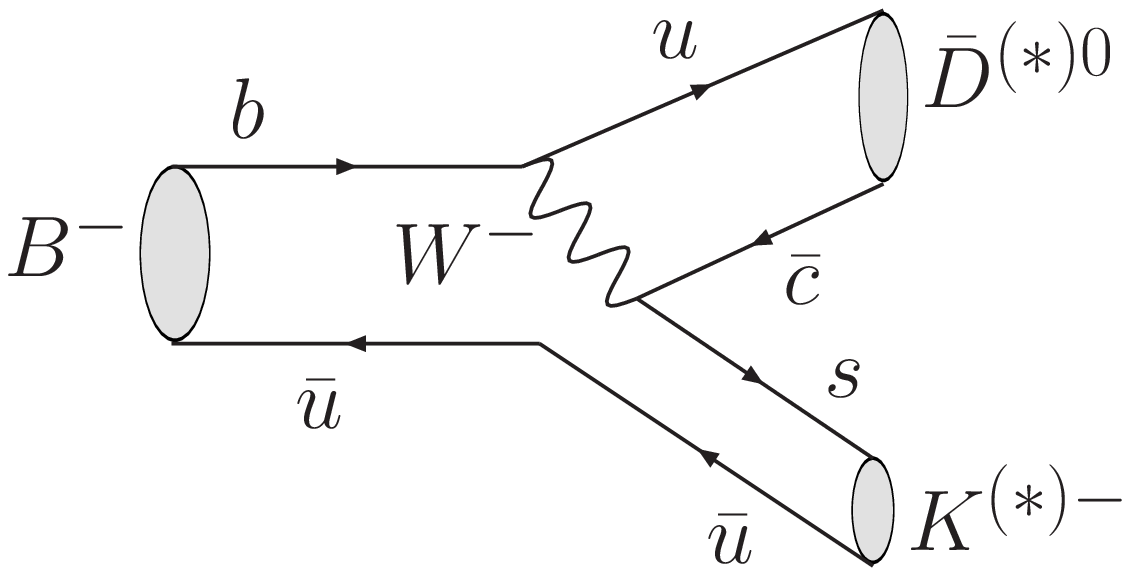,height=3.9cm}
   \caption{Feynman diagrams for $B^- \to D^{(*)0} K^{(*)-}$ and $\Dzbpar K^{(*)-}$.
   The latter is CKM- and color-suppressed with respect to the former.
   The CKM and color suppression factors are expected to be roughly
   $|V_{ub}V_{cs}^*/V_{cb}V_{us}^*| \approx 0.4$ and 1/3 respectively.  }
   \label{fig:feynman}
   \end{center}
   \end{figure}

   Several proposed methods for measuring $\gamma$ exploit the
   interference between $B^- \to D^{(*)0}K^{(*)-}$ 
   and $B^- \to \Dzbpar K^{(*)-}$ 
   (Fig.~\ref{fig:feynman}) which occurs when the $D^{(*)0}$ and the
   \Dzbpar decay to common final states, as first suggested in Ref.~\cite{dk1}.

   As proposed in Ref.~\cite{dk2}, we search for $B^- \to \dbarp~K^-$
   and $B^- \to \dbarpstar~~K^-,~ \dbarpstar~\to \dbarp \pi^0/\gamma$
   followed by $\dbarp~\to K^+\pi^-$, as well as the charge conjugate
   sequences.  In these processes, the favored $B$ decay followed by
   the doubly CKM-suppressed $D$ decay interferes with the suppressed
   $B$ decay followed by the CKM-favored $D$ decay. The interference
   of the $b \to c$ and $b \to u$ amplitudes is sensitive to the
   relative weak phase $\gamma \equiv\arg(-V_{ud}V^*_{ub}/V_{cd}V^*_{cb})$.

   We use the notation $B^- \to [h_1^+h_2^-]_D h_3^-$
   (with each $h_i=\pi$ or $K$) for the decay chain $B^- \to \dbarp~h_3^-$,
   $\dbarp~\to h_1^+ h_2^-$.  For the closely related modes with a 
   $\dbarpstar$~, we use the same notation with the subscript $D$ replaced
   by $D^*$.
   We also refer to $h_3$ as the bachelor
   $\pi$ or $K$.  

   In the decays of interest, the sign of the bachelor kaon is opposite to
   that of the kaon from $D$ decay.  It is convenient to
   define ratios of rates between these decays and the similar decays where
   the two kaons have the same sign.  The decays with same-sign kaons have much higher rate and
   proceed almost exclusively through the
   CKM-favored and color favored $B$ transition, followed by the Cabibbo-favored
   $D$-decay.  The advantage in taking ratios is that most
   theoretical and experimental uncertainties cancel. 
   Thus, ignoring the possible effects of $D$ mixing
   and taking into account the effective strong phase difference of $\pi$ between
   the \Dstar decays in $D \gamma$ and $D \piz$\cite{Bondar}, 
   we define the charge-specific ratios for $D$ and $D^*$ as:

   \begin{equation}
   {\cal R}^{\pm}_{K\pi} \equiv 
   \frac{\Gamma([K^\mp \pi^\pm]_D K^\pm)}{\Gamma([K^\pm \pi^\mp]_D K^\pm)}
   = r_B^2 + r_D^2 + 2 r_B r_D \cos(\pm \gamma + \delta)
   \end{equation}
   \noindent and
   \begin{eqnarray}
   {\cal R}^{*\pm}_{K\pi,D \piz}  \equiv 
   \frac{\Gamma([K^\mp \pi^\pm]_{D^* \ra D\piz} K^\pm)}{\Gamma([K^\pm \pi^\mp]_{D^* \ra D\piz} K^\pm)}
  & =& r_B^{*2} + r_D^2 + 2 r_B^* r_D \cos(\pm \gamma + \delta^*) \ \ \ {\rm and} \\
   {\cal R}^{*\pm}_{K\pi,D\gamma}  \equiv 
   \frac{\Gamma([K^\mp \pi^\pm]_{D^* \ra D\gamma} K^\pm)}{\Gamma([K^\pm \pi^\mp]_{D^* \ra D\gamma} K^\pm)}
  & =& r_B^{*2} + r_D^2 - 2 r_B^* r_D \cos(\pm \gamma + \delta^*),
   \end{eqnarray}
   \noindent where
   \begin{equation}
   r_B \equiv \left| \frac{A(B^- \to \Dzb K^-)}{A(B^- \to D^0 K^-)} \right|, 
   \end{equation}
   \begin{equation}
   r_B^* \equiv \left| \frac{A(B^- \to \Dstarzb K^-)}{A(B^- \to \Dstarz K^-)} \right|, 
   \end{equation}
   \begin{equation}
   \label{eqn:rd}
   r_D \equiv \left| \frac{A(D^0 \to K^+ \pi^-)}{A(D^0 \to K^- \pi^+)} \right|
   = 0.060 \pm 0.003~\cite{dcsdbr},
   \end{equation}
   \begin{equation}
   \delta^{(*)} \equiv \delta_B^{(*)} + \delta_D,
   \end{equation}
   and \noindent $\delta_B^{(*)}$ and $\delta_D$ are 
   strong phase differences between the two $B$ and $D$ decay 
   amplitudes, respectively.

   We also define the charge-integrated ratios:

   \begin{equation}
   {\cal R}_{K\pi} \equiv 
   \frac{\Gamma(B^- \to [K^+ \pi^-]_D K^-)+\Gamma(B^+ \to [K^- \pi^+]_D K^+)}
        {\Gamma(B^- \to [K^- \pi^+]_D K^-)+\Gamma(B^+ \to [K^+ \pi^-]_D K^+)}
   \end{equation}
   \noindent and
   \begin{equation}
   {\cal R}^*_{K\pi,D\piz(D\gamma)} \equiv 
   \frac{\Gamma(B^- \to [K^+ \pi^-]_{D^* \ra D\piz(D\gamma)} K^-)+\Gamma(B^+ \to [K^- \pi^+]_{D^* \ra D\piz(D\gamma)} K^+)}
        {\Gamma(B^- \to [K^- \pi^+]_{D^* \ra D\piz(D\gamma)} K^-)+\Gamma(B^+ \to [K^+ \pi^-]_{D^* \ra D\piz(D\gamma)} K^+)}.
   \end{equation}
   Then,
   \begin{equation}
   {\cal R}_{K\pi} = \frac{{\cal R}^{+}_{K\pi}+{\cal R}^{-}_{K\pi}}{2} =
   r_B^{2} + r_D^2 + 2 r_B r_D \cos\gamma \cos \delta
   \end{equation}
   \begin{equation}
   {\cal R}^{*}_{K\pi,D\piz } = \frac{{\cal R}^{*+}_{K\pi,D \piz}+{\cal R}^{*-}_{K\pi,D \piz}}{2} = 
   r_B^{*2} + r_D^2 + 2 r_B^{*} r_D \cos\gamma \cos \delta^{*}, 
   \label{eq:star-pi}
   \end{equation}
    and
   \begin{equation}
   {\cal R}^{*}_{K\pi,D\gamma} = \frac{{\cal R}^{*+}_{K\pi,D\gamma}+{\cal R}^{*-}_{K\pi,D\gamma}}{2}  = 
   r_B^{*2} + r_D^2 - 2 r_B^{*} r_D \cos\gamma \cos \delta^{*},
   \label{eq:star-gam}
   \end{equation}
   assuming no \CP violation in the normalization modes
   $[K^\mp \pi^\pm]_D K^\mp$ and $[K^\mp \pi^\pm]_{D^*} K^\mp$.
   In the following we use the notation
   $R^*_{K\pi}$ when there is no need to refer specifically to $R^*_{K\pi, D\piz}$
   or $R^*_{K\pi, D\gamma}$.

   Since $r_B^{(*)}$ is expected to be of the same order as $r_D$,
   \CP violation could manifest itself as a large difference between
   the charge-specific ratios
   ${\cal R}^{(*)+}_{K\pi}$ and ${\cal R}^{(*)-}_{K\pi}$.  Measurements of
   these six ratios can be used to constrain $\gamma$.

   The value of $r_B^{(*)}$ determines, in part, the level of interference
   between the diagrams
   of Fig.~\ref{fig:feynman}.  In most techniques for measuring $\gamma$,
   high values of $r_B^{(*)}$ lead to larger interference and better sensitivity to $\gamma$.
   As we will describe below, the measured ${\cal R}^{(*)}_{K\pi}$ are consistent
   with zero in the current analysis.  This allows us to set restrictive upper
   limits on $r_B^{(*)}$, since ${\cal R}^{(*)}_{K\pi}$ depend quadratically on
   $r_B^{(*)}$.

   In the Standard Model,
   $r_B^{(*)} = |V^{}_{ub} V_{cs}^*/V^{}_{cb} V_{us}^*| \: F_{cs} \approx 0.4 \: F_{cs}$.
   The color-suppression factor 
   $F_{cs} < 1$ accounts for the additional suppression, beyond that
   due to CKM factors, of $B^- \to \Dzbpar K^-$ relative to $B^- \to D^{(*)0} K^-$.
   Naively, $F_{cs} = \frac{1}{3}$, which is the probability for the color
   of the quarks from the virtual $W$ in $B^- \to \Dzbpar K^-$ to match 
   that of the other two quarks; see Fig.~\ref{fig:feynman}.  
   Early estimates~\cite{neubert} of $F_{cs}$ were based on factorization and the
   then available experimental information on a number of $b \to c$
   transitions.  These estimates gave $F_{cs} \approx 0.22$, leading to
   $r_B^{(*)} \approx 0.09$.   However, the recent observations and
   measurements~\cite{colorsuppressed} 
   of color suppressed $b \to c$ decays ($B \to D^{(*)} h^0$; 
   $h^0 = \pi^0, \rho^0, \omega, \eta, \eta'$) suggest that $F_{cs}$, and
   therefore $r_{B}^{(*)}$, could be larger.  

   In this paper we report on an update of our previous analysis of 
   $B^- \to \dbarp~~K^-$~\cite{ADS-BABAR}, and the first attempt 
   to study $B^- \to \dbarpstar~~K^-$.  The previous analysis 
   was based on a sample of $B$-meson decays a factor of 1.9 smaller
   than used here,  and resulted in an upper limit ${\cal R}_{K\pi} < 0.026$ at the 90\% 
   confidence level.  This in turn was translated into a limit $r_B < 0.22$,
   also at 90\% C.L..   On the other hand, a study by the Belle collaboration~\cite{belleRB}
   of $B^{\pm} \to \dbarp K^{\pm}$ and
   $B^{\pm} \to \dbarpstar~~K^{\pm}$, $\dbarp \to K_S \pi^+ \pi^-$,
   favors rather large color suppressed amplitudes:  
   $r_B = 0.26^{+0.11}_{-0.15}$ and 
   $r_B^* = 0.20^{+0.20}_{-0.18}$.

\section{THE \babar\ DATASET}
\label{sec:babar}

   The results presented in this paper are based on 
    $227 \times 10^6$ $\FourS\to B\Bbar$ decays,
   corresponding to an integrated luminosity of 205 fb$^{-1}$.
   The data were collected
   between 1999 and 2004 with the \babar\ detector~\cite{babar} at the \pep2\
   \BF\ at SLAC~\cite{pep2}.  
   In addition, a 16~fb$^{-1}$ off-resonance data sample,
   with center-of-mass (CM) 
   energy 40~\mev below the \FourS resonance,
   is used to study backgrounds from
   continuum events, $e^+ e^- \to q \bar{q}$
   ($q=u,d,s,$ or $c$).

\section{ANALYSIS METHOD}
\label{sec:Analysis}
   This work is an extension of our 
   analysis from Ref.~\cite{ADS-BABAR}, which resulted in limits on
   ${\cal R}_{K\pi} < 0.026$ and $r_B < 0.22$, as mentioned above.
   The main changes in the analysis are
   the following:
   \begin{itemize}
   \item The size of the dataset is increased from 
   $120$ to $227 \times 10^6$ $\FourS\to B\Bbar$ decays.
   \item This analysis also includes the $B^{\pm} \to \dbarpstar~~K^{\pm}$ mode.
   \item The analysis requirements have been tightened in order to 
   reduce backgrounds further.
   \item A few of the requirements in the previous analysis resulted
   in small differences in the efficiencies of the signal mode 
   $B^{\pm} \to [K^{\mp}\pi^{\pm}]K^{\pm}$ and the normalization mode
   $B^{\pm} \to [K^{\pm}\pi^{\mp}]K^{\pm}$.  These requirements have now been removed.
   \end{itemize}

   \begin{table}[!htb]
   \caption{Notation used in the text.}
   \begin{center}
   \begin{tabular}{|l|l|l|} \hline
   Abbreviation & Mode & Comments\\ \hline
   $DK$      & $B^- \to D^0 K^-$, $D^0 \to K^- \pi^+$ and c.c. & normalization mode \\
   $D\pi$    & $B^- \to D^0 \pi^-$, $D^0 \to K^- \pi^+$ and c.c. & control mode \\
   $\Dbar K$ & $B^- \to \dbarp K^-$, $\dbarp \to K^+\pi^-$ and c.c. & signal mode \\
   $D^*K$    & $B^- \to D^{*0} K^-$, $D^{*0} \to D^0 \pi^0/\gamma$,  $D^0 \to K^- \pi^+$ and c.c. & normalization mode \\
   $D^*\pi$    & $B^- \to D^{*0} \pi^-$, $D^{*0} \to D^0 \pi^0/\gamma$,  $D^0 \to K^- \pi^+$ and c.c. & control mode\\
   $\Dstarb K$& $B^- \to \dbarpstar~~K^-$, $\dbarpstar~~\to \dbarp \pi^0/\gamma$,
$\dbarp \to K^+ \pi^-$ and c.c. & signal mode \\
   \hline
   \end{tabular}
   \end{center}
   \label{tab:shorthand}
   \end{table}

   \noindent The analysis makes use of several samples from different decay modes.  
   Throughout the following discussion we will refer to these modes 
   using abbreviations that
   are summarized in Table~\ref{tab:shorthand}.

   The event selection is developed from studies of
   simulated $B\Bbar$ and continuum events, and off-resonance
   data. A large on-resonance control sample of $D\pi$ and 
   $D^*\pi$ events 
   is used to validate several aspects of the simulation and 
   analysis procedure.

   The analysis strategy is the following:
   \begin{enumerate}
   \item The goal is to measure or set limits on the
   charge-integrated ratios ${\cal R}_{K\pi}$ and ${\cal R}_{K\pi}^*$. 
   \item The first step consists in the application of a 
   set of basic requirements to select possible candidate events,
   see Section~\ref{sec:basic}.
   \item After the basic requirements, the backgrounds are dominantly from
   continuum.  These are significantly reduced using a neural network designed
   to distinguish between $B\Bbar$ and continuum events,
   see Section~\ref{sec:nn}.
   \item After the neural network requirement, events are characterized by 
   two kinematical variables that are customarily used when reconstructing
   $B$-meson decays at the $\FourS$.  These variables are 
   the energy-substituted mass,
   $\mes \equiv \sqrt{(\frac{s}{2}  + \vec{p}_0\cdot \vec{p}_B)^2/E_0^2 - p_B^2}$
   and energy difference $\Delta E \equiv E_B^*-\frac{1}{2}\sqrt{s}$, 
   where $E$ and $p$ are energy and momentum, the asterisk
   denotes the CM frame, the subscripts $0$ and $B$ refer to the
   \FourS and $B$ candidate, respectively, and $s$ is the square
   of the CM energy.  For signal events $\mes = m_B$ and
   $\Delta E = 0$ within the 
   resolution of about 2.5 and 20 MeV respectively
   (here $m_B$ is the known $B$ mass).
   \item We then perform simultaneous fits to the final signal samples ($\Dbar K$ and $\Dstarb K$),
   the normalization samples ($DK$ and $\Dstar K$),
   and the control samples ($D\pi$ and $\Dstar \pi$)
   to extract ${\cal R}_{K\pi}$ and ${\cal R}_{K\pi}^*$,
   see Section~\ref{sec:fit}. 
   The fits are based on the reconstructed values of $\mes$ and 
   $\Delta E$ in the various event samples.
   \item Throughout the whole analysis chain, care is taken to 
   treat the signal, normalization, and control samples in a consistent manner.
   \end{enumerate}

   \subsection{Basic Requirements}
   \label{sec:basic}
   Charged kaon and pion candidates in the decay modes of interest
   must satisfy $K$ or $\pi$ identification criteria~\cite{PIDref}
   that are typically 85\% efficient, depending on momentum and polar angle.
   The misidentification rates are at the few percent level.
   The invariant mass of the $K\pi$ pair must be within 18.8 MeV
   (2.5$\sigma$) of the mean reconstructed $D^0$ mass.
   For modes with 
   $\dbarpstar~\to \dbarp~\pi^0$ and 
   $\dbarpstar~~\to \dbarp~\gamma$ the mass difference $\Delta M$ between
   the $\dbarpstar~$ and the $\dbarp$ must be within 3.5 
   (3.5$\sigma$) and 13 (2$\sigma$)
   MeV, respectively, of the expectation for $\dbarpstar~$
   decays.
   
   A major background arises from $DK$ and $D^*K$ decays where the 
   $K$ and $\pi$ in the $D$ decay are misidentified as a $\pi$ and
   a $K$ respectively. When this happens, the decay could be reconstructed
   as a $\Dbar K$ or $\Dstarb K$ signal event.
   To eliminate this 
   background, we recompute the invariant mass ($M_{{\rm switch}}$) 
   of the $h^+h^-$ 
   pair in $\dbarp \to h^+h^-$ switching the particle identification
   assumptions ($\pi$ vs. $K$) on the $h^+$ and the $h^-$.  We veto
   any candidates with $M_{{\rm switch}}$ consistent with the known $D$ mass~\cite{PDG}.
   In the case of $\dbarp K$, we also veto any candidate where the
   $\dbarp$ is consistent with $D^*$ decay.

   \subsection{Neural Network}
   \label{sec:nn}
   After these initial requirements, backgrounds are overwhelmingly
   from continuum events, especially $e^+ e^- \to c \bar{c}$, with 
   $\bar{c} \to \Dzb X$, $\Dzb \to K^+ \pi^-$ and $c \to D X$,
   $D \to K^- Y$.

   The continuum background is reduced by using neural network techniques.
   The neural network algorithms used for the $DK$ and $\Dbar K$ modes
   are slightly different.  First, we use for both modes a common neural 
   network ($NN$) based on nine quantities that distinguish between 
   continuum and $B\Bbar$ events.  Then, for the $\Dstarb K$ mode only, 
   we also take advantage of the fact that the signal is distributed
   as $\cos^2\theta_{D^*}$ for $D^* \to D\pi$ or 
   $\sin^2\theta_{D^*}$ for $D^* \to D\gamma$, while the background
   is roughly independent of $\cos\theta_{D^*}$. 
   Here $\theta_{D^*}$ is the decay angle of the $D^*$, {\it i.e.},
   the angle between the direction of the $D$ and the line of flight of
   the $D^*$ relative to the parent $B$, evaluated in the $D^*$ rest
   frame.
   Thus, we
   construct a second neural network, $NN'$, which takes as 
   inputs the output of $NN$ and the value of $\cos\theta_{D^*}$.
   We then use as a selection requirement the output of $NN$ in the 
   $\Dbar K$ analysis and the output of $NN'$ in the $\Dstarb K$ analysis.

   The nine variables used in defining $NN$ are the following:
   \begin{enumerate}
   \item A Fisher discriminant constructed from the 
   quantities $L_0 = \sum_i{p_i}$ and $L_2 = \sum_i{p_i \cos^2\theta_i}$
   calculated in the CM frame.  Here, $p_i$ is the momentum and
   $\theta_i$ is the angle with respect to the thrust axis of the $B$ candidate
   of tracks and clusters not used to reconstruct the $B$ meson.
   \item  $|\cos \theta_T|$, where $\theta_T$ is the angle in
   the CM frame between the thrust axes of the $B$ candidate and
   the detected remainder of the event.  The distribution of 
   $|\cos \theta_T|$ is approximately flat for signal and strongly
   peaked at one for continuum background.
   \item $\cos \theta_B$, where $\theta_B$ is the polar angle
   of the $B$ candidate in the CM frame.  In this variable, the 
   signal follows a $\sin^2 \theta_B$ distribution, while the 
   background is approximately uniform.
   \item $\cos \theta_D^K$ where $\theta_D^K$ is the decay angle
   in $\dbarp \to K\pi$.
   \item $\cos \theta_B^D$, where $\theta_B^D$ is the decay angle
   in $B \to \dbarp K$ or
   $B \to \dbarpstar~~K$.
   \item The charge difference $\Delta Q$ between the sum
   of the charges of tracks in the $\dbarp$ or $\dbarpstar~~$ hemisphere
   and the sum of the charges of the tracks in the opposite
   hemisphere excluding the tracks used in the reconstructed $B$.
   For signal, $\langle \Delta Q \rangle = 0$, whereas
   for the $c\bar{c}$ background 
   $\langle \Delta Q \rangle \approx \frac{7}{3}\times Q_B$,
   where $Q_B$ is the charge of the $B$ candidate.  The
   $\Delta Q$ RMS is 2.4.
   \item $Q_B \cdot Q_K$, where $Q_K$ is the sum of the charges of all
   kaons not in the reconstructed $B$.
   In many signal events, there is a charged kaon among the decay 
   products of the other $B$ in the event.  The charge
   of this kaon tends to be highly correlated with the charge of the 
   $B$.  Thus, signal events tend to have $Q_B \cdot Q_K \leq -1$.  On
   the other hand, most continuum events have no kaons outside of the
   reconstructed $B$, and therefore $Q_K = 0$.
   \item The distance of closest approach between the bachelor
   track and the trajectory of the $\dbarp$.  This is
   consistent with zero for signal events, but can be
   larger in $c\bar{c}$ events.
   \item The existence of a lepton ($e$ or $\mu$) and
   the invariant mass ($m_{K\ell}$) of this lepton and the bachelor $K$.
   Continuum events have fewer leptons than signal events.
   Furthermore, a large fraction of leptons in $c\bar{c}$ events
   are from $D \to K \ell \nu$, where $K$ is the bachelor kaon,
   so that $m_{K\ell} < m_D$.
\end{enumerate}

\begin{figure} \begin{center}
    \epsfig{file=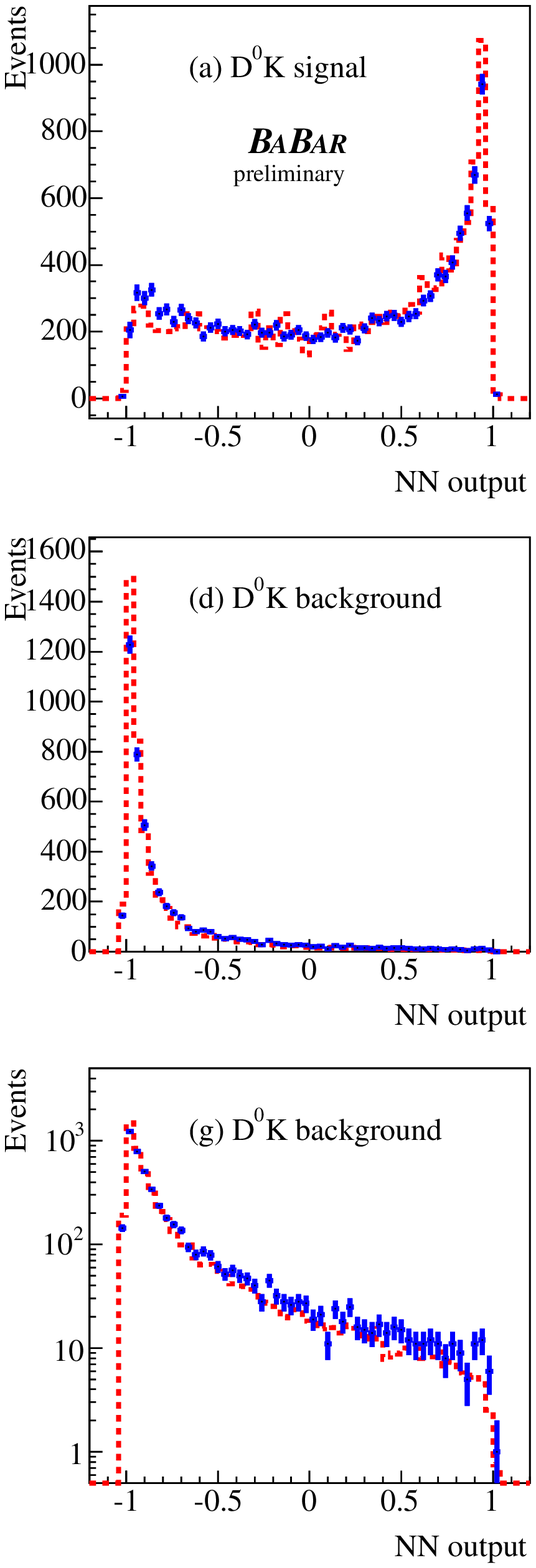,width=0.32\linewidth}
    \epsfig{file=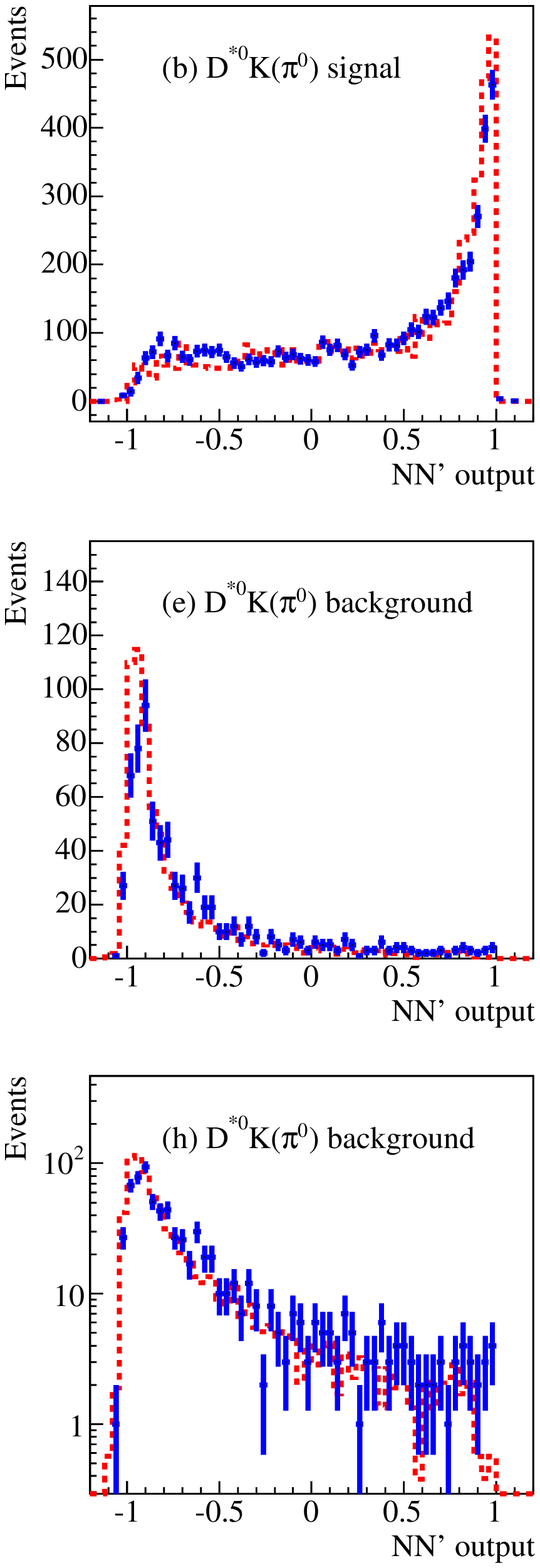,width=0.32\linewidth}
    \epsfig{file=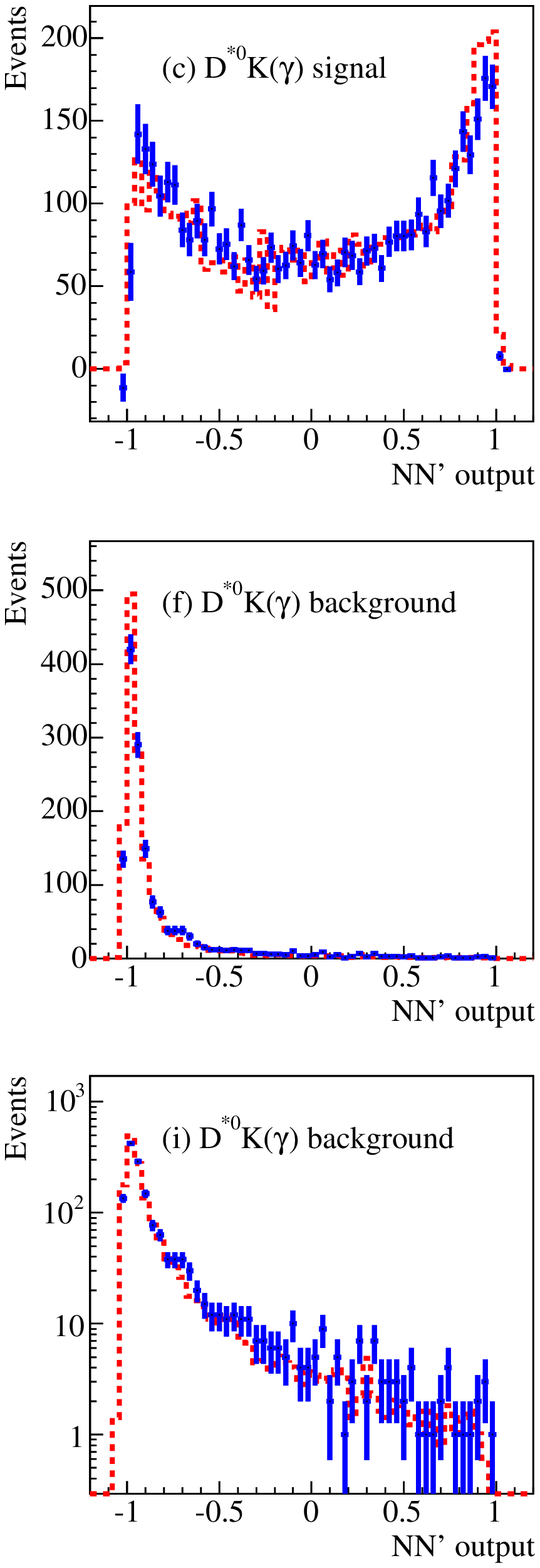,width=0.32\linewidth}
    \caption{ Distributions of the continuum suppression neural
    network ($NN$ and $NN'$) outputs for the three modes. Figures
    (a-c) show the expected distribution from signal events. The solid
    line histogram shows the distribution of simulated signal events,
    the histogram with error bars shows the distribution of
    $D^{(*)0}\pi$ control sample events with background subtracted
    using the $\mes$ sideband. Figures (d-i) show the expected
    distribution for continuum background events. The solid line
    histogram shows the distribution of simulated continuum events and
    the histogram with errors show the distribution of off-resonance
    events. The $\mes$ and $\Delta E$ requirements on the
    off-resonance and continuum Monte Carlo events have been kept
    loose to increase the statistics. Figures (g-i) show the
    distributions of figures (d-g) in log scale.  Each Monte Carlo
    histogram is normalized to the area of the corresponding data
    histogram.}
    \label{fig:nnplots}
    \end{center}
    \end{figure}

   The neural networks ($NN$ and $NN'$) are trained with simulated continuum 
   and signal events.  
   The distributions of the $NN$ and $NN'$ outputs for the control samples  
   ($D\pi$, $D^*\pi$, and off resonance data), are compared with
   expectations from the Monte Carlo simulation
   in Figure~\ref{fig:nnplots}.  The agreement
   is satisfactory.
   We have also examined the distributions of all variables used in $NN$ and $NN'$,
   and found good agreement between the simulation and the data 
   control samples.

   Our final events selection requirement is $NN > 0.5$ for $\Dbar K$ 
   and $NN' > 0.5$ for $\Dstarb K$.  In addition, to reduce the remaining
   $B\Bbar$ backgrounds, we also require
   $\cos \theta_D^K > -0.75$.
   These requirements are about 40\% efficient
   on simulated signal events, and reject 98.5\% of the continuum background.
   Note, however, that we do not rely on the Monte Carlo simulation to estimate
   the efficiency of the neural net requirements.  We apply the exact same
   requirements to the normalization modes $DK$ and $D^{*}K$.  Then, in the
   extraction of ${\cal R}_{K\pi}$ and ${\cal R}_{K\pi}^*$, the 
   efficiencies of the overall selection cancels in the ratio.

   \subsection{Fitting for event yields and 
   {\boldmath ${\cal R^{(*)}_{K\pi}}$}}
   \label{sec:fit}
   The ratios ${\cal R_{K\pi}}$ and ${\cal R^*_{K\pi}}$ are extracted
   from the ratios of the event yields in the $m_{ES}$ distribution
   for the signal modes ($\Dbar K$ and $\Dstarb K$) and the
   normalization modes ($DK$ and $\Dstar K$), while taking into
   account potential differences in efficiencies and backgrounds.  All
   events must satisfy the requirements discussed above and
   have a $\Delta E$ value consistent with zero within the resolution
   ($-52~{\rm MeV} < \Delta E < 44~{\rm MeV}$). 

   The $\mes$ distributions for $\Dbar K$ (signal mode) and $D K$
   (normalization mode) are fitted simultaneously.  
   The fit parameter ${\cal R}_{K\pi}$ is given by 
   ${\cal R}_{K\pi} \equiv c \cdot N_{\Dbar K}/N_{DK}$,
   where $N_{\Dbar K}$ and $N_{DK}$ are the fitted yields of
   $\Dbar K$ and $DK$ events, and $c$ is a correction factor, determined
   from Monte Carlo, for the ratio of efficiencies between the two 
   modes.  We find that this factor $c$ is consistent with unity
   within the statistical accuracy of the simulation,
   $c = 0.98 \pm 0.04$\footnote{In the $D^*$ modes this correction factor is 
   $c = 0.97 \pm 0.05$ and $c = 0.99 \pm 0.05$ for $D^* \to D \pi^0$ 
   and $D^* \to D \gamma$ respectively.}.

   The $\mes$ distributions are modeled as the sum of a threshold
   combinatorial background function~\cite{ARGUS} and a Gaussian
   centered at $m_B$.  The parameters of the background function for
   the signal mode are constrained by a simultaneous fit of the $\mes$
   distribution for events in the sideband of $\Delta E$ 
   ($-120~{\rm MeV} < \Delta E < 200~{\rm MeV}$, excluding the
   $\Delta E$ signal region defined above).
   The parameters of the Gaussian for the signal
   and normalization modes are constrained to be identical. The number
   of events in the Gaussian is $N_{sig} + N_{peak}$, where $N_{sig} =
   N_{DK}$ or $N_{\Dbar K}$ and $N_{peak}$ is the number of background
   events expected to be distributed in the same way as the $DK$ or
   $\Dbar K$ in $\mes$ (``peaking backgrounds'').

   There are two classes of peaking background events:

   \begin{enumerate}  
   \item Charmless $B$ decays, {\em e.g.}, $B^- \to K^+ K^- \pi^+$.
   These are indistinguishable from the $\Dbar K$ signal if the $K^-\pi^+$
   pair happens to be consistent with the $D$-mass.   
   \item Events
   of the type $B^- \to D^0 \pi^-$ ($D\pi$), where the bachelor $\pi^-$
   is misidentified as a $K^-$.  When the $D^0$ decays into
   $K^-\pi^+$ ($K^+\pi^-$), these events are indistinguishable
   in $\mes$ from $DK$ ($\Dbar K$), since $\mes$ is insensitive 
   to particle identification assumptions.
   \end{enumerate}

   The amount of charmless background (1) is estimated directly from the
   data by performing a simultaneous fit to events in the sideband of
   the reconstructed $D$ mass.
   The $\Delta E$ distribution of the $D\pi$ background (2) is 
   shifted by about $+ 50$ MeV due to the
   misidentification of the bachelor $\pi$ as a $K$. Since the
   $\Delta E$ resolution is of order 20 MeV, the $\Delta E$
   requirement does not eliminate this
   background completely.  The remaining $D\pi$ background after the
   $\Delta E$ requirement is estimated from a fit to
   the $\Delta E$ distribution of the $DK$ sample.

   We fit the $\Delta E$ distribution of $DK$ candidates, with $\mes$
   within $3 \sigma$ of $m_B$, to the sum of a $DK$ component, a
   $D\pi$ component, and a combinatorial component.  The $D\pi$
   sample, with the bachelor track identified as a pion, is used to
   constrain the shape of the $DK$ component in the $DK$ sample.  The
   same sample of $D\pi$ events, but
   reconstructed as $DK$ events, is used to
   constrain the shape of the $D\pi$ background in the $DK$ sample.
   The fitted number of $D\pi$ background events in this sample that
   survive the $\Delta E$ requirements, which we denote as
   $N^{\pi}_{DK}$, is taken as the number of $D\pi$ background events
   in the fit to the $m_{ES}$ distribution of $DK$ events..

   The $D\pi$ peaking background is much more
   important in the $DK$ (normalization) channel than in the $\Dbar K$
   (signal) channel.  This is because in order to contribute to the
   signal channel, the $D^0$ has to decay into $K^+ \pi^-$, and this
   mode is doubly Cabibbo-suppressed.
   For the $\Dbar K$ (signal) sample, the contribution from the residual 
   $D\pi$ peaking background in the $\mes$ fit is estimated as 
   $N^{\pi}_{\Dbar K} = r^2_D N^{\pi}_{DK}$, where $r_D = 0.060 \pm 0.003$
   is the ratio of the doubly Cabibbo-suppressed to the Cabibbo-favored
   $D \to K\pi$ amplitudes (see Eq.~\ref{eqn:rd}), and $N^{\pi}_{DK}$ was defined above.

   The complete procedure simultaneously fits seven distributions: the
   $m_{ES}$ distributions of $DK$ and $\Dbar K$, the $\Dbar K$
   distributions in sidebands of $\Delta E$ and $m(D^0)$, the $\Delta
   E$ distribution of $DK$, and the $\Delta E$ distributions of $D \pi$
   reconstructed as $D \pi$ and as $DK$.  The fits are configured in
   such a way that ${\cal R_{K\pi}}$ and ${\cal R^*_{K\pi}}$ are
   explicit fit parameters.  The advantage of this approach is that
   all uncertainties, including the uncertainties in the PDFs and the
   uncertainties in the background subtractions, are automatically
   correctly propagated in the statistical uncertainty reported by the
   fit. 

   The fit is performed separately for $\Dbar K$, $\Dstarb K$,
   $\Dstarb \to \Dbar \pi^0$, and $\Dstarb K$, $\Dstarb \to \Dbar
   \gamma$ and is identical for all three modes, except in the choice of
   parameterization for some signal and background components in the
   $\Delta E$ fits.   

   Systematic uncertainties in the detector
   efficiency cancel in the ratio.  
   This cancellation has been verified by studies of simulated events,
   with a statistical precision of a few per-cent.  
   The likelihood includes a Gaussian uncertainty term for this cancellation
   which is set by the statistical accuracy of the simulation.  Other
   systematic uncertainties, e.g., the uncertainty in the parameter
   $r_D$ which is used to estimate the amount of peaking backgrounds
   from $D^{(*)}\pi$, are also included in the formulation of the
   likelihood.

   The fit procedure has been extensively tested on sets
   of simulated events. It was found to provide an unbiased estimation
   of the parameters ${\cal R_{K\pi}}$ and ${\cal R^*_{K\pi}}$.

\section{RESULTS}

\begin{figure}[hbt]
    \begin{center}
    \epsfig{file=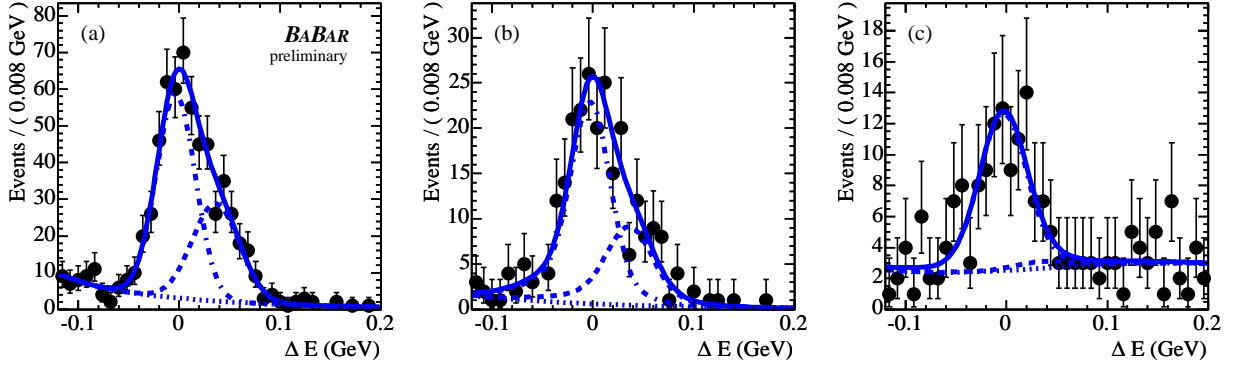,width=\linewidth}
    \caption{ $\Delta E$ distributions for normalization events
    ($DK$ and $D^*K$) with $\mes$ within 3$\sigma$ of $m_{B}$
    with the fit model overlaid.  
    (a) $DK$ events.
    (b) $D^*K$ events with $D^* \to D\pi^0$. 
    (c) $D^*K$ events with $D^* \to D\gamma$.
    The dashed (dot-dashed) curves are the contributions from $D\pi$ or $D^*\pi$
    ($DK$ or $D^*K$) 
    events.  The dotted curves are the contributions from other
    backgrounds, and the solid line is the total.}
    \label{fig:de_all}
    \end{center}
  \end{figure}

\begin{figure}[hbt]
    \begin{center}
    \epsfig{file=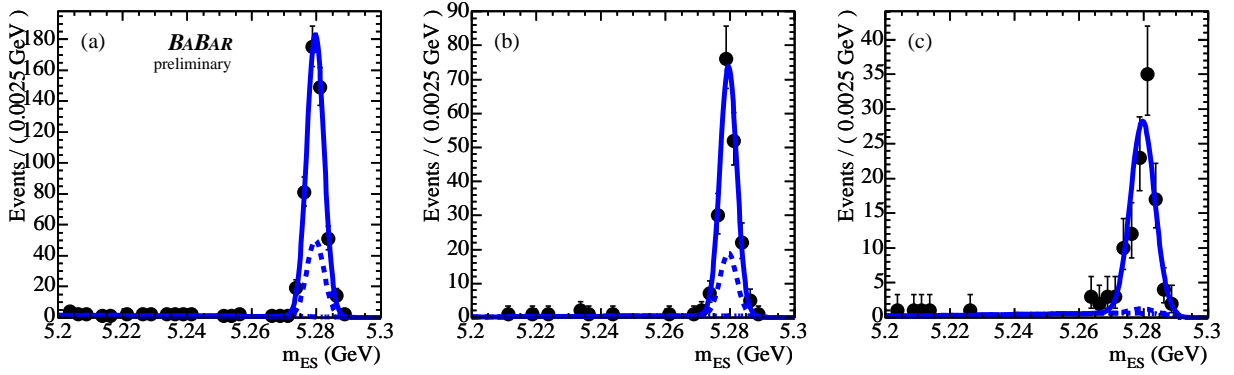,width=\linewidth}
    \caption{\protect $\mes$ distributions for normalization events
    ($DK$ and $D^*K$) with $\Delta E$ in the signal region
    with the fit model overlaid.  (a) $DK$ events.
    (b) $D^*K$ events with $D^* \to D\pi^0$. 
    (c) $D^*K$ events with $D^* \to D\gamma$.
    The dashed curves represent the backgrounds; these are mostly
    from $D\pi$ or $D^*\pi$, and also peak at the $B$-mass.  
    As explained in the text, the size of the $D\pi$ and
    $D^*\pi$ backgrounds is constrained by the simultaneous 
    fits to the distributions of Figure~\ref{fig:de_all}.}
    \label{fig:mes_d0k_all}
    \end{center}
  \end{figure}

\begin{figure}[hbt]
    \begin{center}
    \epsfig{file=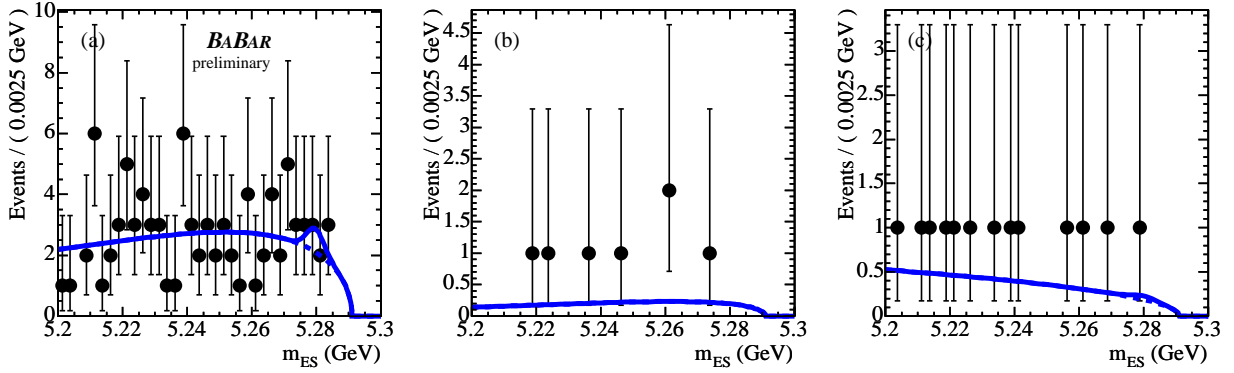,width=\linewidth}
    \caption{ $\mes$ distributions for $\Dbar K$ and $\Dstarb K$
    events with $K\pi$ mass in a sideband of the reconstructed 
    $D$ mass and
    with $\Delta E$ in the signal region.  These events are used 
    to constrain the size of possible peaking backgrounds from
    charmless
    $B$-meson decays, {\em i.e.}, decays without a $D$ in the final state.
    The fit model is overlaid.  (a) $\Dbar K$ events.
    (b) $\Dstarb K$ events with $D^* \to D\pi^0$. 
    (c) $\Dstarb K$ events with $D^* \to D\gamma$.
    Note that the $K\pi$ mass range in the sideband selection
    is a factor of 2.7 larger than in the signal selection.}
    \label{fig:mes_d0sb_all}
    \end{center}
    \end{figure}

\begin{figure}[hbt]
    \begin{center}
    \epsfig{file=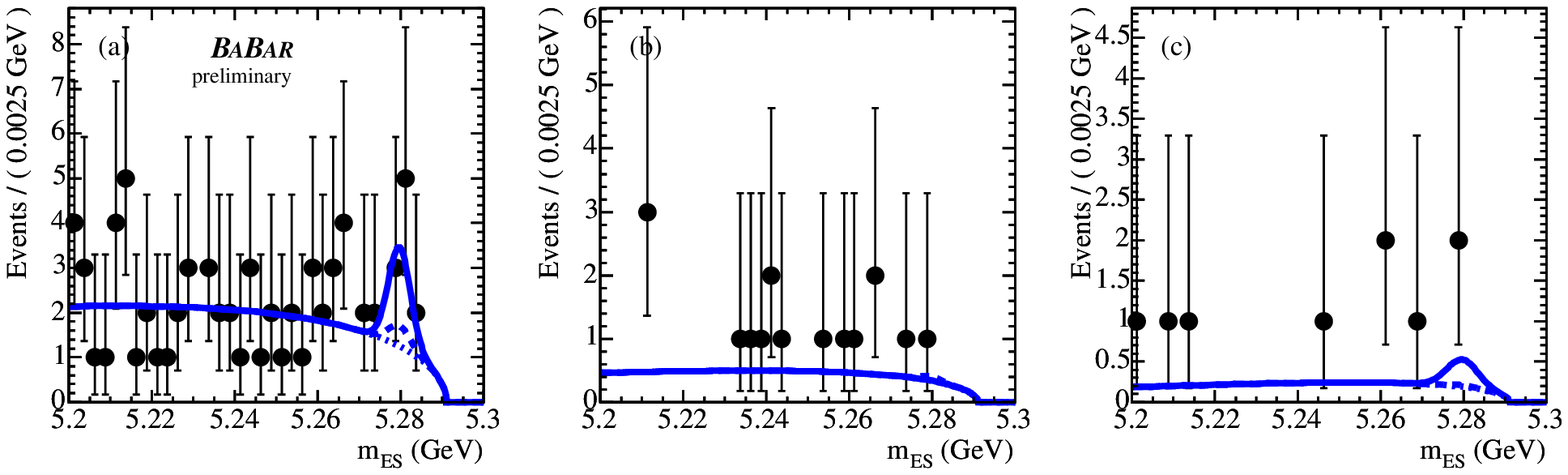,width=\linewidth}
    \caption{ $\mes$ distributions for candidate signal
    events
    with the fit model overlaid.  (a) $\Dbar K$ events.
    (b) $\Dstarb K$ events with $D^* \to D\pi^0$. 
    (c) $\Dstarb K$ events with $D^* \to D\gamma$.}
    \label{fig:mes_sig_all}
    \end{center}
    \end{figure}

  The results of the fits are displayed in Table~\ref{tab:fit-results}
  and
  Figs.~\ref{fig:de_all},~\ref{fig:mes_d0k_all},~\ref{fig:mes_d0sb_all},
  and~\ref{fig:mes_sig_all}.  As is apparent from
  Fig.~\ref{fig:mes_sig_all}, we see no evidence for the $\Dstarb K$
  modes and no significant evidence for the $\Dbar K$ mode.

  For the $\Dbar K$ mode we find ${\cal R_{K\pi}} = 13^{+11}_{-9}
  \times 10^{-3}$; For the $\Dstarb K$ mode we find ${\cal R^*_{K\pi}}
  = -1^{+10}_{-6} \times 10^{-3}$ (for $D^* \to D \pi^0$) and ${\cal
  R^*_{K\pi}} = 11^{+19}_{-13} \times 10^{-3}$ (for $D^* \to D
  \gamma$).  
  We estimate from a
  parameterized Monte Carlo study that the probability that an upward
  fluctuation of background events results in our observed value of 
  ${\cal R_{K\pi}}$ or larger is 7.5\%.

   \begin{table}[htb]
   \caption{Summary of fit results.}
   \begin{center}
   \begin{tabular}{|l|l|l|l|} \hline
        &          &  & \\
   Mode & $\Dbar K$  &  $\Dstarb K$, $\Dstarb \to \Dbar \pi^0$ & $\Dstarb K$, $\Dstarb \to \Dbar \gamma$ \\ \hline 
   Ratio of rates, ${\cal R_{K\pi}}$ or ${\cal R^*_{K\pi}}$, $\times 10^{-3}$&
   ${\cal R_{K\pi}} = 13^{+11}_{-9}$ &  
   ${\cal R^*_{K\pi}} = -1^{+10}_{-6}$ &  
   ${\cal R^*_{K\pi}} = 11^{+19}_{-13}$ \\
   No. of signal events & $4.7^{+4.0}_{-3.2}$ & $-0.2^{+1.3}_{-0.8}$ & $1.2^{+2.1}_{-1.4}$ \\
   No. of normalization events & $356 \pm 26$ & $142 \pm 17$ & $101 \pm 14$ \\ 
   No. of peaking charmless events                 & $0.75^{+1.3}_{-0.75}$   & $0^{+0.3}_{-0.0}$ & $0.05^{+0.7}_{0.05}$ \\
   No. of peaking $D^{(*)}\pi$ ev. in sig. sample  & $0.47 \pm 0.04$ & $0.17 \pm 0.02$ & $0.01^{+0.03}_{-0}$ \\
   No. of peaking $D^{(*)}\pi$ ev. in norm. sample & $132 \pm 10$  & $48 \pm 6$       & $2.5 \pm 8$\\
   \hline
   \end{tabular}
   \end{center}
   \label{tab:fit-results}
   \end{table}

\begin{figure}[hbt]
    \begin{center}
    \epsfig{file=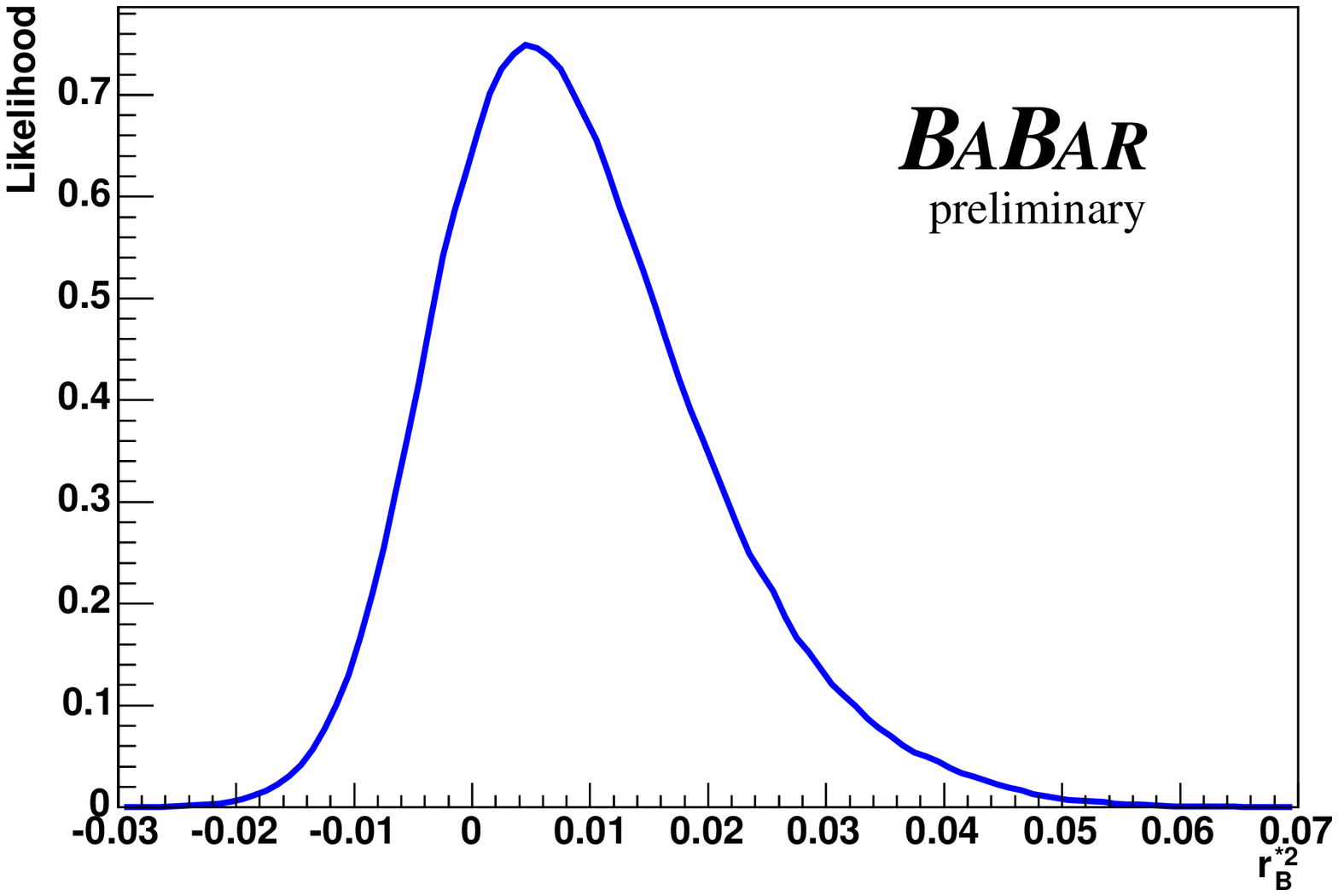,width=10cm}
    \caption{ Likelihood distribution (arbitrary units)
    for $r_B^{*2}$.}
    \label{fig:likelihood2}
    \end{center}
    \end{figure}

  We use our results to extract information on $r_B$ and $r^*_B$.
  In the case of decays into a $\Dstar/\Dstarb$ we can use equations~\ref{eq:star-pi} and~\ref{eq:star-gam}
  to write
   \begin{equation}
   r_B^{*2} = 
   \frac{{\cal R}^{*}_{K\pi,D \piz}+{\cal R}^{*}_{K\pi,D\gamma}}{2} - r_D^2.
   \end{equation}
   Our results then give $r_B^{*2} = 4.6^{+15.2}_{-7.3} \times 10^{-3}$.  
  The likelihood function for 
  $r_B^{*2}$ is shown in Figure~\ref{fig:likelihood2}.  Based on this likelihood,
  we set an upper limit $r_B^{*2} < (0.16)^2$ at the 90\% C.L. using a Baysean method
  with a uniform prior for $r_B^{*2} > 0$.

    \begin{figure}[hbt]
    \begin{center}
    \epsfig{file=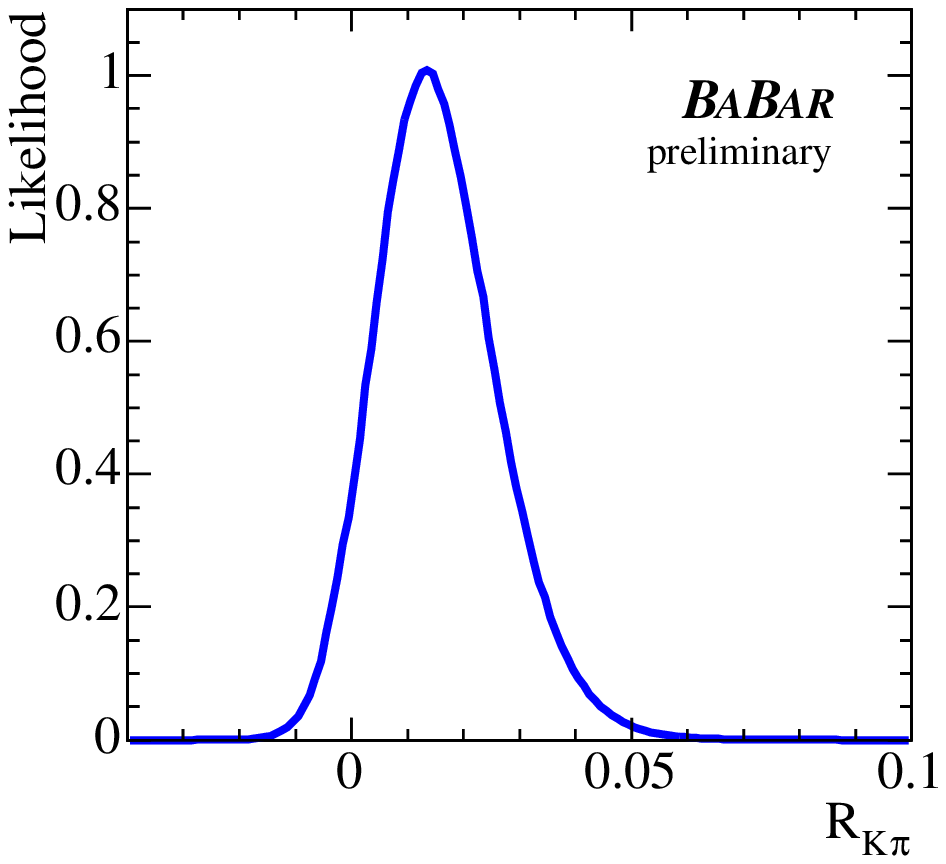,width=7cm}
    \caption{ Likelihood distribution (arbitrary units)
    for ${\cal R_{K\pi}}$.}
    \label{fig:likelihood}
    \end{center}
    \end{figure}

\begin{figure}[htb]
   \begin{center}
   \epsfig{file=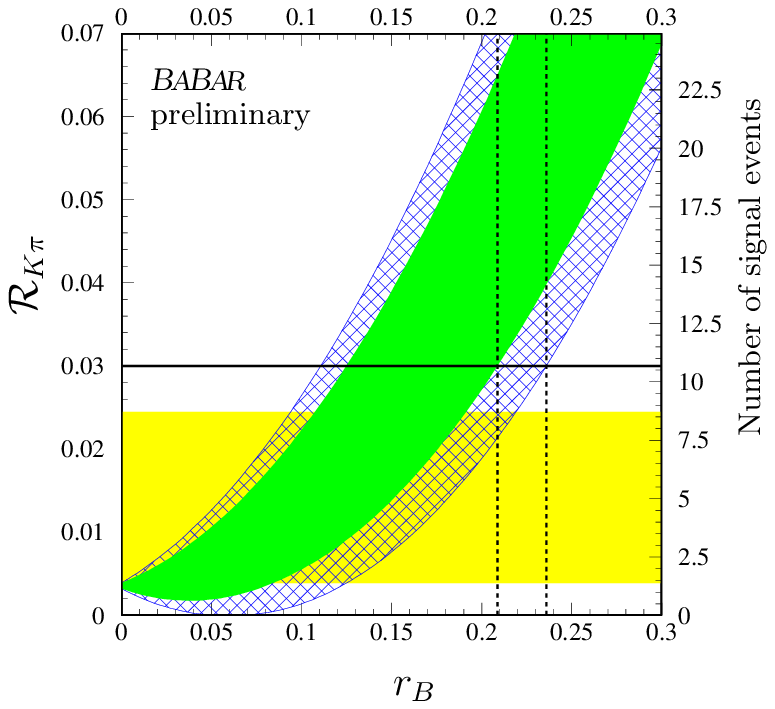,width=0.49\linewidth}
   \caption{Expectations for ${\cal R}_{K\pi}$ and the number
   of signal events {\it vs.} $r_B$.  Dark filled-in area:
   allowed region for any value of $\delta$, with a $\pm
   1\sigma$ variation on $r_D$, and $48^{\circ} < \gamma <
   73^{\circ}$.  Hatched area: additional allowed regions with no
   constraint on $\gamma$.  Note that the uncertainty on $r_D$ has a
   very small effect on the size of the allowed regions.  The
   horizontal line represents the 90\% C.L.  limit ${\cal R}_{K\pi} <
   0.030$. The vertical dashed lines
   are drawn at $r_B = 0.209$, $r_B = 0.235$.  
   They represent the 90\% C.L. upper limits on
   $r_B$ with and without the constraint on $\gamma$. The
   light filled in areas represent the 68\% C.L. regions corresponding
   to ${\cal R}_{K\pi} = 0.013 \pm^{0.011}_{0.009}$. }
   \label{fig:ads_rate}
   \end{center}
   \end{figure}

  In the case of decays into a $D/\Dbar$, there are not enough information to 
  extract the ratio $r_B$ without additional assumptions.  Thus, we first extract an 
  upper limit on the experimentally measured
  quantity ${\cal R_{K\pi}}$ .  This is done starting 
  from the likelihood as a function of ${\cal R_{K\pi}}$ 
  (see Figure~\ref{fig:likelihood})
  using a Bayesian method
  with a uniform prior for ${\cal R_{K\pi}} > 0$.
  The limit is ${\cal R_{K\pi}} < 0.030$
  at 90\% C.L..
   Next, in Fig.~\ref{fig:ads_rate} we show the dependence of
   ${\cal R}_{K\pi}$ on $r_B$, together with our limit on 
   ${\cal R}_{K\pi}$.
   This dependence is shown allowing a $\pm 1\sigma$ variation on $r_D$,
   for the full range $0^{\circ}-180^{\circ}$ for $\gamma$ and
   $\delta$, as well as with the restriction
   $48^{\circ} < \gamma < 73^{\circ}$ suggested by global CKM
   fits~\cite{ckmfitter}.  We use the information displayed
   in this Figure to set an upper limit on $r_B$. 
   The least restrictive limit on $r_B$
   is computed assuming maximal destructive interference:
   $\gamma=0^{\circ}, \delta = 180^{\circ}$ or 
   $\gamma=180^{\circ}, \delta = 0^{\circ}$.  
   The limit is $r_B < 0.23$ at 90\% C.L..

\clearpage

\section{SUMMARY}
\label{sec:Summary}

   In summary, we find no significant evidence for the decays $B^{\pm}
   \to [K^{\mp}\pi^{\pm}]_D K^{\pm}$ and 
   $B^{\pm} \to [K^{\mp}\pi^{\pm}]_{D^*} K^{\pm}$.  
   We measure the ratios ${\cal R}_{K\pi}$ of the rates for these modes 
   and the favored modes $B^{\pm} \to [K^{\pm}\pi^{\mp}]_D K^{\pm}$ 
   and $B^{\pm} \to [K^{\pm}\pi^{\mp}]_{D^*} K^{\pm}$ as
   ${\cal R}_{K\pi} = 0.013 \pm^{0.011}_{0.009}$, 
   ${\cal R}^*_{K\pi,D\piz} = -0.001 \pm^{0.010}_{0.006}$ and 
   ${\cal R}^*_{K\pi,D\gamma} = 0.011 \pm^{0.019}_{0.013}$.

   We use the results for ${\cal R}^*_{K\pi,D\piz}$ and  
   ${\cal R}^*_{K\pi,D\gamma}$ to set a model independent limit 
   $r^{*2}_B \equiv |A(B^- \to \Dzb K^-)/A(B^- \to \Dz K^-)|^2 < (0.16)^2$ 
   at the 90\% confidence level. 
   We also set 90\%
   C.L. limits on the ratio ${\cal R}_{K\pi} < 0.030$ (90\% C.L.). 
   With the most conservative assumption on the
   values of $\gamma$ and of the strong phases in the $B$ and $D$
   decays, this limit translates into a limit on the ratio of the magnitudes
   of the $B^- \to \Dzb K^-$ and $B^- \to \Dz K^-$ amplitudes
   amplitude
   $r_B < 0.23$ at 90\% C.L..  If $r_B$ and $r_B^*$ are
   small, as our analysis suggests, the suppression of the
   $b\rightarrow u$ amplitude will make the determination of $\gamma$
   using methods based on the interference of the diagrams in
   Figure~\ref{fig:feynman} difficult.

\section{ACKNOWLEDGMENTS}
\label{sec:Acknowledgments}

We are grateful for the 
extraordinary contributions of our \pep2\ colleagues in
achieving the excellent luminosity and machine conditions
that have made this work possible.
The success of this project also relies critically on the 
expertise and dedication of the computing organizations that 
support \babar.
The collaborating institutions wish to thank 
SLAC for its support and the kind hospitality extended to them. 
This work is supported by the
US Department of Energy
and National Science Foundation, the
Natural Sciences and Engineering Research Council (Canada),
Institute of High Energy Physics (China), the
Commissariat \`a l'Energie Atomique and
Institut National de Physique Nucl\'eaire et de Physique des Particules
(France), the
Bundesministerium f\"ur Bildung und Forschung and
Deutsche Forschungsgemeinschaft
(Germany), the
Istituto Nazionale di Fisica Nucleare (Italy),
the Foundation for Fundamental Research on Matter (The Netherlands),
the Research Council of Norway, the
Ministry of Science and Technology of the Russian Federation, and the
Particle Physics and Astronomy Research Council (United Kingdom). 
Individuals have received support from 
CONACyT (Mexico),
the A. P. Sloan Foundation, 
the Research Corporation,
and the Alexander von Humboldt Foundation.

\end{document}